\documentstyle[12pt,equations,doublespace,cite]{article}
\def\bold#1{\setbox0=\hbox{$#1$}%
     \kern-.025em\copy0\kern-\wd0
     \kern.05em\copy0\kern-\wd0
     \kern-.025em\raise.0433em\box0 }

\def\slash#1{\setbox0=\hbox{$#1$}#1\hskip-\wd0\dimen0=5pt\advance
       \dimen0 by-\ht0\advance\dimen0 by\dp0\lower0.5\dimen0\hbox
         to\wd0{\hss\sl/\/\hss}}
\newcommand{\dd}{\displaystyle}
\newcommand{\nn}{\nonumber}
\newcommand{\smallz}{{\scriptscriptstyle Z}} %  a smaller Z
\newcommand{\smallw}{{\scriptscriptstyle W}} %
\newcommand{\smallh}{{\scriptscriptstyle H}} %
\newcommand{\shat}{ {\hat s} }
\newcommand{\chat}{ {\hat c} }
\newcommand{\acur}{ {\hat \alpha} }
\newcommand{\mz}{M_{\smallz}}
\newcommand{\mw}{M_{\smallw}}
\newcommand{\mh}{M_{\smallh}}
\newcommand{\mt}{m_t}
\newcommand{\mzb}{M_{\smallz_0}}
\newcommand{\mwb}{M_{\smallw_0}}

\newcommand{\mtb}{m_{t_0}}
\newcommand{\sineff}{\mbox{$\sin^2 \theta^{{\rm lep}}_{{\rm eff}}$} }
\newcommand{\kl}{\mbox{$\hat{k}_l$}}
\newcommand{\scur}{\mbox{$\hat{s}^2$}}
\newcommand{\ccur}{\mbox{$\hat{c}^2$}}

\newcommand{\drhoc}{\mbox{$ \delta \rho_{\scriptscriptstyle{C}}$}}
\newcommand{\dlr}{\mbox{$ \Delta r$}}
\newcommand{\drcar}{\mbox{$\Delta \hat{r}$}}
\newcommand{\dro}[1]{\mbox{$ \delta \rho^{{\scriptscriptstyle (#1)}}$}}
\newcommand{\drof}{\mbox{$ \delta \rho^{{\scriptscriptstyle f(1)}}$}}
\newcommand{\drob}{\mbox{$ \delta \rho^{{\scriptscriptstyle b(1)}}$}}

\newcommand{\pz}[1]{\mbox{$ \Pi_{\smallz \smallz}(#1) $}}
\newcommand{\azz}{\mbox{$ A_{\smallz \smallz} $}}
\newcommand{\aww}{\mbox{$ A_{\smallw \smallw} $}}
\newcommand{\Azz}[1]{\mbox{$ A_{\smallz \smallz}^{
{\scriptscriptstyle #1}}$}}
\newcommand{\Aww}[1]{\mbox{$ A_{\smallw \smallw}^{
{\scriptscriptstyle #1}}$}}
\newcommand{\gmu}{\mbox{$ G_\mu $}}
\newcommand{\gmud}{\mbox{$ O(G_\mu^2 m_t^2 \mz^2) $}}
\newcommand{\gmuq}{\mbox{$ O(G_\mu^2 m_t^4) $}}
\newcommand{\selfs}{self-energies}
\newcommand{\self}{self-energy}

\newcommand{\ew}{electroweak}
\newcommand{\msbar}{\overline{MS}}

\newcommand{\Vw}{\mbox{$V_\smallw$}}
\newcommand{\Bw}{\mbox{$B_\smallw$}}
\newcommand{\Vz}{\mbox{$V_\smallz$}}
\newcommand{\Bz}{\mbox{$B_\smallz$}}

\newcommand{\equ}[1]{Eq.~(\ref{#1})}
\newcommand{\eqs}[1]{Eqs.~(\ref{#1})}
\newcommand{\Eqs}[2]{Eqs.~(\ref{#1}) and (\ref{#2})}
\newcommand{\efe}[1]{Ref.\cite{#1}}

\newcommand{\be}{\begin{equation}}
\newcommand{\ee}{\end{equation}}
\newcommand{\een}{\end{subequations}}
\newcommand{\ben}{\begin{subequations}}
\newcommand{\beq}{\begin{eqalignno}}
\newcommand{\eeq}{\end{eqalignno}}
\newcommand{\beqtwo}{\begin{eqaligntwo}}
\newcommand{\eeqtwo}{\end{eqaligntwo}}
\newcommand{\beqs}{\begin{eqalignno*}}
\newcommand{\eeqs}{\end{eqalignno*}}
\newcommand{\bea}{\begin{eqnarray}}
\newcommand{\eea}{\end{eqnarray}}

\begin{document}
\begin{titlepage}

\title{\bf Two-loop electroweak top corrections: \\
           are they under control? $^*$ }

\author{
G.~Degrassi$^{~a}$,~S.~Fanchiotti$^{~b}$,
{}~F.~Feruglio$^{~a}$,B
P.~Gambino$^{~c}$,~A.~Vicini$^{~a}$\\
}
\date{}
\maketitle

\vspace{1cm}

\begin{itemize}
\item[$^a$]

Dipartimento di Fisica, Universit\`a  di Padova,
Sezione INFN di Padova \\
Via Marzolo 8, 35131 Padova, Italy \\[-10mm]

\item[$^b$]
Theory Division, CERN, CH-1211 Geneva 23, Switzerland\\[-10mm]

\item[$^c$]
Department of Physics, New York University, New York,
NY 10003, USA \\[-10mm]

\end{itemize}

\vspace{1cm}

\begin{abstract}
The assumption that two-loop top corrections are well approximated
by the $\gmuq$ contribution is investigated. It is shown that in the case
of the ratio neutral-to-charged current amplitudes at zero momentum
transfer the $\gmud$ terms are numerically comparable to the
$m_t^4$ contribution for realistic values of the top mass.
An estimate of the theoretical error due to unknown two-loop
top effect is presented for a few observables of LEP interest.
\end{abstract}

\vfill

\noindent
emails:\\
degrassi@mvxpd5.pd.infn.it \\[-2mm]
sergio@mafalda.physics.nyu.edu \\[-2mm]
feruglio@ipdgr4.pd.infn.it,\\[-2mm]
gambino@acf2.nyu.edu \\[-2mm]
vicini@mvxpd5.pd.infn.it \\

\vfill

\footnoterule

\noindent
$^*${\footnotesize
To appear in "Reports of the Working Group on Precision
Calculations for the $Z$-resonance", CERN.
This research is partially supported by EU under contract
No.~CHRX-CT92-0004.}

\end{titlepage}
\section{Introduction}
The constant improvement of the experimental precision
on line shape and asymmetry parameters at LEP has stimulated
the evaluation of two-loop  corrections of a purely electroweak
nature in order
to assess the reliability of the theoretical predictions.
Although the latter seem to be affected mainly  by the uncertainty
of the hadronic contribution on $\Delta\alpha$, it is not yet clear which
error may be
attributed to the ignorance of higher orders in the electroweak
perturbative expansion. The first attempt made in this direction was the
computation of the Higgs contribution to the $\rho$  parameter in the
limit of large
$\mh$ \cite{VvB}. Subsequently, top effects were also investigated
\cite{vBH}. Concerning the top, we only have at the moment
two-loop results obtained from the
SM in the limit of vanishing gauge coupling constants \cite{bar,FTJ,DHL}.
Such  contributions are of $\gmuq$ and formally leading in the limit of
large top mass. They should be considered as the present best estimate
of the top influence on higher-order corrections.
This note deals with the next-to-leading corrections of
$\gmud$. Such terms are suppressed by a power $\mz^2/\mt^2$ with
respect to the leading ones, but the present range of values for
$\mt$ \cite{CDF,Lep}
does not exclude that these corrections may be numerically important.
Our computation can be regarded as an attempt to check
the validity of such an expansion, until the full two-loop
results are available. At the same time we should be able
to provide a more realistic estimate
of the error associated with the two-loop electroweak effects.

To keep the computation as simple as possible we have focused
on neutrino scattering on a leptonic target, of which we
will compute the electroweak
corrections of $\gmud$ to the $\rho$ parameter, defined as the ratio
of neutral-to-charged current amplitudes, at zero momentum transfer.
To be more precise,  we identify $\rho$ with the cofactor, expressed in
units of $G_\mu$, the $\mu$-decay constant, of the $J_\smallz \,
J_\smallz$ interaction
in neutral current amplitudes. It is well known that radiative effects
also lead to a
modification of the mixing angle, described by a parameter usually called
$\kappa$. These effects will not be discussed in the present paper.

For the processes under examination, we  found  large subleading
corrections of the same sign and of about the same magnitude as
the leading one. Therefore, at least for the case we have investigated,
the use of the first term of an expansion in inverse power of $\mt$
to approximate the full two-loop result appears to be doubtful.
Our result, being obtained at $q^2=0$, cannot be directly
applied to LEP physics, but can give us a flavour of the size of
subleading effects that are due to one-particle irreducible
contributions.
In the concluding Section, we will elaborate this point,
analysing the consequences
of a na\"\i ve extrapolation of our result to some LEP observables.
\section{\gmud\ corrections to the $\rho$ parameter.}
In this Section we outline the computation of the electroweak
corrections of $\gmud$ to the $\rho$ parameter.
We begin by writing the relation between the $\mu$-decay constant and
the
charged current amplitude expressed in terms of bare quantities. At the
two-loop level, neglecting contributions that will not give \gmud ~terms,
we have
\be
\frac{\gmu}{\sqrt2} = \frac{g_0^2}{8 \mwb^2} \left\{
1-\dd\frac{\aww}{\mwb^2} + \Vw + \mwb^2 \Bw +
\dd\frac{\aww^2}{\mw^4}-\dd\frac{\aww\Vw}{\mw^2}\right\}~~~,
\label{e2.1}
\ee
where $g_0$ and $\mwb$ are  the bare $SU(2)_L$ coupling and $W$ mass,
respectively, $\aww$ is the transverse part of the $W$ \self\ at zero
momentum
transfer, and  the quantities $\Vw$ and $\Bw$ represent the relevant
vertex and box corrections. At the bare level, using the fact that
$\mzb^2
c_0^2= \mwb^2$, where $c_0 \equiv \cos \theta_{\smallw_{0}}$ with
$\theta_{\smallw_{0}}$ the weak mixing angle and $\mzb$ the bare
 $Z$ mass, the $\rho$ parameter can be written as:
\be
\rho=
\frac
{\left(1-\dd\frac{\azz}{\mzb^2} + \Vz + \mzb^2 c_0^2 \Bz +
\dd \frac{\azz^2}{\mz^4} -\dd\frac{\azz\Vz}{\mz^2}\right)}
{\left(1-\dd\frac{\aww}{\mwb^2} + \Vw + \mwb^2 \Bw +
\dd\frac{\aww^2}{\mw^4}-\dd\frac{\aww\Vw}{\mw^2}\right)}~~~,
\label{e2.2}
\ee
where $\azz,\, \Vz$ and $\Bz$ are the corresponding \self, vertex, and
box contribution in the neutral current amplitude.
To the order we are interested in, \equ{e2.2} reduces to:
\bea
\rho=1&+&\left(\frac{\aww}{\mwb^2}-\frac{\azz}{\mzb^2}\right)
      +(\Vz-\Vw)+ (\mwb^2+\aww)(\Bz-\Bw)\nn\\[1.5mm]
     &+&\left(\frac{\aww}{\mw^2}-\frac{\azz}{\mz^2}\right)
       \left(-\frac{\azz}{\mz^2}+(\Vz-\Vw) - \mw^2 \Bw \right)~~~.
\label{e2.2bis}
\eea
We proceed by separating the self-energies into one-loop
and two-loop contributions:
\beqtwo
\azz&=\Azz {(1)} +\Azz {(2)}~;&~~~\aww&=\Aww {(1)} +\Aww {(2)}~~~,
\label{e2.3}
\eeqtwo
on the understanding that the one-loop term is still expressed in terms
of bare parameters.
The one-loop part can be decomposed further into pure bosonic ($b$) and
fermionic ($f$) terms:
\beqtwo
\Azz {(1)}&=\Azz {b (1)}+\Azz {f (1)}~;&
{}~~~\Aww {(1)}&=\Aww {b (1)}+\Aww {f (1)}~~~,
\label{e2.4}
\eeqtwo
and the one-loop fermionic contribution to the $\rho$ parameter,
assuming a vanishing bottom mass, can be expressed as follows:
\ben \label{e2.5} \beq
X^0_d&=\left(\frac{\Aww {f}}{\mwb^2}-\frac{\Azz {f}}{\mzb^2}\right)^{
(1)} =
\frac{g_0^2}{8 \mwb^2} f(\mtb^2,\epsilon)
\label{e2.5a}\\
f(\mt^2,\epsilon)& \equiv \frac{3}{2 \pi^2}~\dd\frac{1}{(4-2\epsilon)}~
\mt^2~\epsilon~
\Gamma(\epsilon) \left(\dd\frac{4 \pi \mu^2}{\mt^2}\right)^{
\dd\epsilon}~~~.
\label{e2.5b}
\eeq \een
where $\epsilon$ is related to the dimension $d$ of the space--time by
$\epsilon = (4 -d)/2$ and $\mu$ is the 't-Hooft mass scale.

We want to express our final result in terms of the physical $Z$ mass,
therefore we perform the shift $\mzb^2 = \mz^2-{\rm Re}~\pz {\mz^2}$,
where $\pz{\mz^2}$ is the transverse part of the $Z$ self-energy at
$q^2\!=\!\mz^2$.
Using the decompositions given in \Eqs{e2.3}{e2.4},
and keeping only terms up to $\gmud$, we obtain after simple algebra:
\bea
\rho&=&1+X^0_d + X_d \left(-\dd\frac{\aww}{\mw^2}+\Vw+
       \mw^2 \Bw\right)\nn\\[1.5mm]
&+&\left(\frac{\Aww {b}/c_0^2 -\Azz {b}}{\mz^2}\right)^{(1)}+
\left(\frac{\aww}{\mw^2}-\frac{\azz}{\mz^2}\right)^{(2)}\nn\\[1.5mm]
&+&(\Vz-\Vw)+ \mz^2 c_0^2 (\Bz-\Bw) -X_d(\Vw+2~ \mw^2 \Bw)\nn\\[1.5mm]
&+&X_d \left[\left(\dd\frac{\aww}{\mw^2}-\dd\frac{\azz}{\mz^2}\right)
+(\Vz-\Vw)+ \mw^2 (\Bz-\Bw)\right]~~~,
\label{e2.6}
\eea
where $X_d$ is the same quantity introduced in Eq.\ (\ref{e2.5}),
but expressed
in terms of renormalized parameters.

We observe that \equ{e2.6} further simplifies if we express
the one-loop fermionic contribution
in terms of the Fermi constant $G_\mu$. Indeed, as  can be seen from
 Eq.\ (\ref{e2.1}), the first line of Eq.\ (\ref{e2.6})
reproduces the effective coupling in the charged sector:
\bea
X^0_d \left(1-\dd\frac{\aww}{\mw^2}+\Vw+\mw^2 \Bw\right)&=&
\dd\frac{g_0^2}{8 \mwb^2}\left(1-\dd\frac{\aww}{\mw^2}+\Vw+\mw^2
\Bw\right)
f(\mtb^2,\epsilon)\nn\\[1mm]
&\simeq & \frac{G_\mu}{\sqrt{2}}~f(\mtb^2,\epsilon)~~~.
\label{e2.7}
\eea
Until now, apart from the use of the physical $Z$ mass, we have
 not specified any particular renormalization condition.
In order to simplify the structure of the counterterms,
we have found it convenient to perform the calculation using the
$\msbar$ parameter $\sin ^2 {\hat \theta}_{W} (\mz)$ (henceforth
abbreviated as $\scur$). Indeed, while in the on-shell (OS) scheme
the counterterm associated with the quantity $s^2=1-\mw^2/\mz^2$
contains terms proportional to $\mt^2$ and gives rise to $\gmud$
contributions to $\rho$, the counterterm related to $\scur$
does not exhibit any $\mt^2$ dependence and this greatly simplifies
our task. Therefore, to the order we are interested in, we can
directly replace
$c_0^2$ with \ccur\ in \equ{e2.6} ($\ccur \equiv 1 - \scur$).
It will always be possible to recover the result
in the pure OS scheme, by appropriately shifting $\scur$ in
the one-loop expression for $\rho$.

We now notice that the one-loop contribution is still written in terms
of bare quantities. To put $\rho$ in its final form, we split it
into the usual $O({\alpha})$ result, $\dro {1}$,
plus the counterterm part, $\drhoc$, namely
\be
\frac{G_\mu}{\sqrt{2}}~f(\mtb^2,\epsilon)+
\left(\frac{\Aww {b}/\ccur-\Azz {b}}{\mz^2}\right)^{(1)}+
(\Vz-\Vw)^{(1)}+ \mz^2 \ccur (\Bz-\Bw)^{(1)} \equiv \dro {1} +\drhoc
\label{e2.8}
\ee
with
\ben \label{e2.9} \beq
\dro {1} &=\drof +\drob \label{e2.9a}\\
\drof& =N_c x_t \equiv N_c \dd\frac{ G_\mu \mt^2}{8 \pi^2 \sqrt{2}}
\label{e2.9b}\\
\drob&= { {\acur}\over {4 \pi \shat^2} }
          \left[ {3\over{4\shat^2} } \ln\chat^2 - {7 \over 4} +
             {{2~ c_Z}\over {\chat^2}} + \shat^2~G(\xi ,\chat^2)
 \right]~~~,
\label{e2.9c}
\eeq \een
where $N_c$ is the colour factor, and
$\acur = \alpha / (1 + 2 \delta e/e)_{\scriptscriptstyle{\msbar}}$
is the $\msbar$ coupling as defined in \cite{DFS}. In \eqs{e2.9}
\ben \label{e2.10} \beq
c_Z &= {{\chat^2}\over 4} (5-3I_3) - 3 \left( {{I_3}\over 8} -
          {{\shat^2}\over 2} Q + \shat^4~I_3~ Q^2 \right)~~~,
\label{e2.10a}
\eeq
where $I_3$ and $Q$ are the isospin and electric charge of the target
($I_3 = -1$ for electrons) and
\beq
G(\xi,\chat^2) &= {3\over 4} {{\xi}\over{\shat^2}}
     \left[ { {\ln\chat^2 -\ln\xi}\over{\chat^2-\xi} } + {1\over
{\chat^2}}
           {{\ln\xi}\over{1-\xi}}\right]~~~, \label{e2.10b}
\eeq \een
with $\xi \equiv \mh^2 / \mz^2$.
Using eqs.~(\ref{e2.6}), (\ref{e2.7}), and (\ref{e2.8}) we can express
$\rho$ as follows:
\be
\rho = 1+\dro {1} + N_c x_t \dro {1} + \dro {2}~~~,
\label{e2.11}
\ee
where the previous relation defines the two-loop contribution,
\dro{2}, as:
\beq
\dro {2}&=\drhoc
+\left(\frac{\aww}{\mw^2}-\frac{\azz}{\mz^2}\right)^{(2)}
+ (\Vz-\Vw)^{(2)}+ \mz^2 \ccur (\Bz-\Bw)^{(2)}\nn\\
&~~~~~-X_d(\Vw+2~\mw^2 \Bw)
\label{e2.12}
\eeq
Eq.~(\ref{e2.11}) suggests that a possible way to take into account
higher-order effects is to write $\rho$ as
\be
\rho = \dd\frac{1}{(1-\drof)}(1+\drob + \dro {2})~~~, \label{e2.13}
\ee
where the resummation of \drof\ can be justified theoretically  on
the basis
of $1/N_c$ expansion arguments \cite{Pe}.
Explicitly we find, in units $N_c\, [\acur/(16 \pi \scur \ccur)
\, \mt^2 / \mz^2]^2 \simeq N_c x_t^2$:
\ben \label{e2.14} \bea
\dro {2}&=
& 25 - 4\,{\it ht} + \left( {1\over 2} - {1\over {{\it ht}}} \right) \,
    {{\pi }^2} + {{\left( -4 + {\it ht} \right) \,{\sqrt{{\it ht}}}\,
       g({\it ht})}\over 2} + \left( -6 - 6\,{\it ht} +
       {{{{{\it ht}}^2}}\over 2} \right) \,\ln {\it ht}\nn\\ [2mm]
  &&+ \left( -15 + {6\over {{\it ht}}} +
      12\,{\it ht} - 3\,{{{\it ht}}^2} \right) \,{\it Li_2}(1 - {\it ht})
     + \left( -15 + 9\,{\it ht} -
     {{3\,{{{\it ht}}^2}}\over 2} \right) \,
     \phi \left({{{\it ht}}\over 4}\right) \nn\\[3mm]
  &&+~ {\it zt}~\,\left[ {{25}\over 2}\right. + {4\over {{\it ht}}} -
      10\,{{\hat c}^2} + {3\over {{{\hat s}^2}}} +
      {{277\,{{\hat s}^2}}\over 9} -
      {{4\,{{\hat s}^2}}\over {{\it ht}}}  \\[2mm]
  &&+ \left( 9 + {3\over {{{\hat s}^4}}} - {6\over {{{\hat s}^2}}} -
       6\,{{\hat s}^2} \right) \,
       \ln {{\hat c}^2} + 3\,\left( 5 - 6\,{{\hat s}^2} \right)
       \,\ln {\it zt} ~+ 6\,{\it I_3}\,{{\hat c}^2} \nn\\[2mm]
   &&+ \left( 2 - {4\over {{\it ht}}} -
    8\,{{\hat s}^2} +
     {{28\,{{\hat s}^2}}\over {{\it ht}}} \right) \,\ln {\it ht}
    + {{\pi }^2}\,\left( -{7\over 3} - {2\over {3\,{{{\it ht}}^2}}} +
         {1\over {{\it ht}}} - {{56\,{{\hat s}^2}}\over {27}} +
         {{2\,{{\hat s}^2}}\over {3\,{{{\it ht}}^2}}} -
     {{{{\hat s}^2}}\over {{\it ht}}}   \right) \nn\\[2mm]
  && + {{12\,\left( -4 +
     {\it ht} \right) \,{{\hat s}^2}\,}\over {{\it ht}}}
      {\it \Lambda}\left(-1 + {4\over {{\it ht}}}\right)  +
         \left(2~ {\it ht}\,~  {{\hat c}^2}  -
         {{2\,\left( -2 + 3\,{\it ht} \right) \,
             {{\hat c}^2} }\over {{{{\it ht}}^2}}} \right) \,
       {\it Li_2}(1 - {\it ht}) \nn\\[2mm]
  &&\left.  + \left( -2 - {8\over {{\it ht}}} + 5\,{{\hat s}^2} +
         {{24\,{{\hat s}^2}}\over {{{{\it ht}}^2}}} -
         {{10\,{{\hat s}^2}}\over {{\it ht}}} + {\it ht}\, {{\hat c}^2}
          \right) \,\phi \left({{{\it ht}}\over 4}\right) \right],
\label{e2.14a}
\eea
for $\mh \gg \mz$, whilst in the region $\mh \ll \mz$,
\beq
\dro {2} &= 19 - 2\,{{\pi }^2} - 4\,\pi\,{\sqrt{{\it ht}}}  +
   {\it ht}\,\left( -{{27}\over 2} + 2\,{{\pi }^2} - 6\,\ln {\it ht} -
      5\,\ln {{\hat c}^2} + 3\,\ln {\it zt} \right) \nn \\
   &~~~~ + {\it zt}\,\left[ -{{11}\over 2} + {3\over {{{\hat s}^2}}} +
      {{319\,{{\hat s}^2}}\over 9} + 6~{\it I_3}~{{\hat c}^2}  +
      {{\pi }^2}\,\left( -{7\over 3} - {{56\,{{\hat s}^2}}\over {27}}
      \right)\right.\nn \\
  &~~~~\left.+ \left( 7 + {3\over {{{\hat s}^4}}} -
     {6\over {{{\hat s}^2}}}
     - 4\,{{\hat s}^2} \right) \,
   \ln {{\hat c}^2} + \left( 21 - 16\,{{\hat s}^2} \right)
   \,\ln {\it zt} \right].
\label{e2.14b}
\eeq
\een
In \eqs{e2.14} $ht \equiv (\mh/\mt)^2$, $zt \equiv (\mz/\mt)^2$,
\ben \label{e2.15} \beq
g(x) =&  \left\{
          \begin{array}{lr}
           \sqrt{4-x}\,\left
	      (\pi - 2 \arcsin{\sqrt{x/4}}
	               \right) & 			0 < x \leq 4 \\
	   {}\\
	   2 \sqrt{x/4-1}\,\ln\left(
	   	\frac{1-\sqrt{1-4/x}}{1+\sqrt{1-4/x}}
		              \right) & 		x > 4 \,~ ,
	  \end{array}
	\right.  \label{e2.15a}
\eeq
\beq
\Lambda(-1 + {4 \over x}) =&  \left\{
          \begin{array}{lr}
          -\frac{1}{2 \sqrt{x}}\,g(x) + {\pi \over 2} \,\sqrt{4/x -1}
 	              & 			0 < x \leq 4 \\
	   {}\\
            -\frac{1}{2 \sqrt{x}}\, g(x)& 		x > 4 \,~ ,
	  \end{array}
	\right.  \label{e2.15abis}
\eeq
\beq
Li_2 (x) =& - \int_0^x dt {\ln (1-t) \over t}  ~~~ ,   \label{e2.15b}
\eeq
and
\be
\phi(z) =
       \begin{cases}
       4 \sqrt{{z \over 1-z}} ~Cl_2 ( 2 \arcsin \sqrt z ) & $ 0 < z
\leq 1$\\
       { 1 \over \lambda} \left[ - 4 Li_2 ({1-\lambda \over 2}) +
       2 \ln^2 ({1-\lambda \over 2}) - \ln^2 (4z) +\pi^2/3 \right]
       & $z >1 $\,,
       \end{cases}
       \label{e2.15c}
\ee
where $Cl_2(x)= {\rm Im} \,Li_2 (e^{ix})$ is the Clausen function
with
\be
\lambda = \sqrt{1 - {1 \over z}}~. \label{e2.15d}
\ee \een
The first two lines of eq.\ (\ref{e2.14a}) represent the leading $\gmuq$
result \cite{bar}, which is completely independent of the gauge
sector of the theory. Indeed this part can be computed in the
framework of a pure Yukawa theory, obtained from the SM in the limit
of vanishing gauge coupling constants.
The rest of eq.\ (\ref{e2.14a}) is proportional to $zt=\mz^2/\mt^2$
and represents the first correction to the Yukawa limit.
Eqs.~(\ref{e2.14}) show a process-dependent contribution, i.e.\ $6\,
zt\, I_3\,
\ccur$ that comes from $\Bz^{(2)}$.
This reflects the fact that, already at one-loop, the
box diagrams in neutral current depend on the process under consideration
\cite{MS80} [cf.\ \equ{e2.10a}].
\section{Numerical results}

In the previous Section we derived the expression for the $\rho$
parameter
up to \gmud\ in the $\msbar$ scheme. We expressed our result in terms
of the
$\msbar$ quantities $\acur$, \scur, and the physical mass of the $Z$
boson.
To obtain the corresponding expressions in terms of \gmu\ and the
on-shell
(OS) parameter $c^2 \equiv \mw^2 / \mz^2$~, we use the relations
\cite{DFS}
\ben \label{e3.1} \beq
\frac{\acur}{4 \pi \scur} & = \frac{\gmu \mw^2}{2 \sqrt{2} \pi^2}~
\frac{1 - \Delta \hat{r}_\smallw}{1 + (\frac{2 \delta e}{e})_{\msbar}}
\simeq \frac{\gmu \mz^2 c^2}{2 \sqrt{2} \pi^2} \label{e3.1a}\\
\ccur & = c^2 (1 - Y_{\overline{MS}}) \simeq c^2 ( 1 - N_c x_t)~~~.
\label{e3.1b}
\eeq
\een

Eq.~(\ref{e3.1b}) will create additional contributions to \dro{2}.
The one-loop
result is then given by \eqs{e2.9}\ with the substitutions
$\acur/ (4 \pi \scur) \rightarrow (\gmu \mz^2 c^2)/ (2 \sqrt{2} \pi^2),
{}~~~~ \scur,\, \ccur \rightarrow s^2,\,c^2$, while for the two-loop
contribution
we have
%\begin{figure}[t]
%\begin{center}
%\mbox{
%\epsfig{file=fig1.ps,height=11.5cm,width=12cm,
%bbllx=99pt,bblly=166pt,bburx=499pt,bbury=586pt}
%            }
%\caption{\dro{2} for $\nu_\mu\, e$ scattering, in units $N_c x_t^2$
%as a function of $\mt$ for few values
% of $\mh$: including only the \gmuq\    contribution $(y)$, and with
% both the
% \gmuq\ and \gmud\ terms $(g)$.}
%\end{center}
%\end{figure}
\ben \label{e3.2} \beq
\delta \rho^{{\scriptscriptstyle (2)}}_{{\scriptscriptstyle OS}}  &=
 \dro{2} (\scur,\,\ccur \rightarrow s^2,\,c^2) \nn \\
& + N_c x_t^2 zt \left[
  - {3 c^4 \over s^4} \ln c^2 - {3 c^2 \over s^2} -3 I_3 + 12 Q
 - 24 s^2 (1 + c^2) I_3 Q^2 + 4 c^2 G^\prime (\xi,c^2) \right]~~~~
\label{e3.2a}
\eeq
where
\be
G^\prime (\xi,c^2) = {3\over 4}~ \xi
     \left[c^2 {\ln (c^2/\xi)\over (c^2-\xi)^2}  - {1\over c^2- \xi}
         +\frac1{c^2}  {{\ln\xi}\over{1-\xi}}\right]~~~. \label{e3.2b}
\ee \een

In \equ{e3.2a}\ $ \dro{2} (\scur,\,\ccur \rightarrow s^2,\,c^2)$
represents
a term obtained from \eqs{e2.14} applying the same substitutions
as in the one-loop case.

\ From \equ{e3.2a} we notice that the process-dependence is more
pronounced in
the OS framework. This is easily understood by noticing that the
expansion of
the bare couplings in the one-loop box diagrams gives rise, unlike
the $\msbar$ case, to $\mt^2$ contributions.

\begin{table}[t]
\begin{center}
{\bf Table 1} \\
\end{center}
$\dro{2}$ ($\msbar$) and
$\delta \rho^{{\scriptscriptstyle (2)}}_{{\scriptscriptstyle OS}}$
($OS$)
relevant to $\nu_\mu\,e$ scattering for $zt \equiv \mz^2/\mt^2 =
0.2,0.3$,
in units $N_c x_t^2$ as
a function of $r=\mh/\mt$.
The column $zt=0$ is the result of the Yukawa theory.\\[-3mm]
\begin{center}
\begin{tabular}{|c|c|c|c|c|c|}
\hline
\multicolumn{1}{|c}{$~ $} &\multicolumn{1}{c|}{$~ $} &
\multicolumn{2}{c|}{$~$} & \multicolumn{2}{c|}{$~$}\\ [-3.5mm]
\multicolumn{1}{|c}{$~ $} &\multicolumn{1}{c|}{$~ $} &
\multicolumn{2}{c|}{$\msbar$} & \multicolumn{2}{c|}{$OS$}\\ \hline
\multicolumn{1}{|c|}{$~ $} &\multicolumn{1}{c|}{$~ $} &
\multicolumn{1}{c|}{$~ $} &\multicolumn{1}{c|}{$~ $} &
\multicolumn{1}{c|}{$~$} & \multicolumn{1}{c|}{$~$}\\ [-3.5mm]
\multicolumn{1}{|c|}{$\displaystyle r={{\mh}\over{\mt}}$}
&\multicolumn{1}{c|}{$ zt~=~0 $} & \multicolumn{1}{c|}{$zt~=~0.2$} &
\multicolumn{1}{c|}{$zt~=~0.3$} & \multicolumn{1}{c|}{$zt~=~0.2$} &
\multicolumn{1}{c|}{$zt~=~0.3$} \\
\multicolumn{1}{|c|}{$~ $} &\multicolumn{1}{c|}{$~ $} &
\multicolumn{1}{c|}{$~ $} &\multicolumn{1}{c|}{$~ $} &
\multicolumn{1}{c|}{$~$} & \multicolumn{1}{c|}{$~$}\\ [-3.5mm]
\hline
\hline
0.1 & $-~1.8$ & --12.6 & --15.8 & --12.7 & --16.0 \\[-1.7mm]
0.2 & $-~2.7$ & --13.3 & --16.5 & --13.5 & --16.8 \\ [-1.7mm]
0.3 & $-~3.5$ & --13.9 & --17.0 & --14.2 & --17.4 \\ [-1.7mm]
0.4 & $-~4.1$ & --14.5 & --17.6 & --14.9 & --18.1 \\ [-1.7mm]
0.5 & $-~4.7$ & --15.2 & --18.3 & --15.7 & --18.9 \\ [-1.7mm]
0.6 & $-~5.2$ & --16.1 & --20.2 & --16.7 & --20.9 \\ [-1.7mm]
0.7 & $-~5.7$ & --16.2 & --20.1 & --16.9 & --20.9 \\ [-1.7mm]
0.8 & $-~6.2$ & --16.4 & --20.1 & --17.1 & --21.0 \\ [-1.7mm]
0.9 & $-~6.6$ & --16.5 & --20.1 & --17.4 & --21.2 \\ [-1.7mm]
1.0 & $-~6.9$ & --16.6 & --20.1 & --17.6 & --21.3 \\ [-1.7mm]
1.1 & $-~7.3$ & --16.8 & --20.2 & --17.8 & --21.4 \\ [-1.7mm]
1.2 & $-~7.6$ & --16.9 & --20.2 & --18.0 & --21.6 \\ [-1.7mm]
1.3 & $-~7.9$ & --17.0 & --20.2 & --18.2 & --21.7 \\ [-1.7mm]
1.4 & $-~8.2$ & --17.2 & --20.3 & --18.4 & --21.9 \\ [-1.7mm]
1.5 & $-~8.4$ & --17.3 & --20.3 & --18.6 & --22.0 \\ [-1.7mm]
1.6 & $-~8.7$ & --17.4 & --20 4 & --18.7 & --22.1 \\ [-1.7mm]
1.7 & $-~8.9$ & --17.5 & --20.5 & --18.9 & --22.3 \\ [-1.7mm]
1.8 & $-~9.1$ & --17.6 & --20.5 & --19.1 & --22.4 \\ [-1.7mm]
1.9 & $-~9.3$ & --17.7 & --20.6 & --19.2 & --22.6 \\ [-1.7mm]
2.0 & $-~9.5$ & --17.8 & --20.6 & --19.4 & --22.7 \\ [-1.7mm]
2.5 & $-10.2$ & --18.2 & --20.9 & --20.0 & --23.3 \\ [-1.7mm]
3.0 & $-10.8$ & --18.4 & --20.8 & --20.4 & --23.5 \\ [-1.7mm]
3.5 & $-11.2$ & --18.3 & --20.6 & --20.6 & --23.6 \\ [-1.7mm]
4.0 & $-11.4$ & --18.3 & --20.4 & --20.6 & --23.5 \\ [-1.7mm]
4.5 & $-11.6$ & --18.2 & --20.1 & --20.6 & --23.4 \\ [-1.7mm]
5.0 & $-11.7$ & --18.0 & --19.8 & --20.5 & --23.3 \\ [-1.7mm]
5.5 & $-11.8$ & --17.8 & --19.4 & --20.4 & --23.1 \\ [-1.7mm]
6.0 & $-11.8$ & --17.5 & --19.0 & --20.3 & --22.9 \\
\hline
\end{tabular}
\end{center}
\end{table}

In Fig.~1 we plot $\dro{2}$ [\eqs{e2.14}] as a function of $\mt$ for few
values of $\mh$. As a comparison we also show the values obtained
including
only the $\gmuq$ contribution. The process under consideration
is $\nu_\mu\, e$
scattering. From Figure 1 it is evident that the inclusion of corrections
suppressed by a factor $\mz^2 /\mt^2$ with respect to the leading term
 is quite substantial.

To have a better understanding of the size of these corrections
in Table 1 we present the values of $\dro{2}$ and
$\delta \rho^{{\scriptscriptstyle (2)}}_{{\scriptscriptstyle OS}} $ for
$zt=0, 0.2$,~and $0.3$ as a function of $r=\mh/\mt$.
When preparing the Table we
matched the values from (\ref{e2.14a}) and (\ref{e2.14b}) when the
latter were very close ($r \simeq 0.5$). We see that in the
region of light Higgs the $\gmud$ corrections are much larger than
the $\mt^4$ term that is actually suppressed by accidental
cancellations, while for large Higgs mass,
in the TeV region, their contribution is still $50$\%  of the leading
part.
It is worth noticing that the numbers shown in Table 1 are very close
to the corresponding ones obtained in \efe{DFG} in the case of a model
with $SU(2)$
symmetry. That is not surprising $\shat$ being  a relatively small number
($\scur \simeq 0.23$).

\section{Conclusions}
 We have seen that the calculation of the
difference of \selfs\ is not sufficient to compute the \gmud\
corrections to
the $\rho$  parameter [cf.\ \equ{e2.12}]
but one has to resort to physical processes and this introduces
process-dependent quantities. Our result, being obtained at $q^2=0$,
cannot
be directly applied to LEP physics. However one can ask general questions
about the two-loop \ew\ corrections involving the top and use the answers
coming from the calculation of $\dro{2}$ as a ``ringing bell'' for the
estimation
of the theoretical error in the present knowledge of these corrections.

It is natural to ask whether we can expect that the \gmuq\
term will approximate well the complete unknown result for values of
$\mt$
not larger than $250$ GeV. Table 1 shows that in the case of $\dro{2}$
the  answer is
negative. We have looked for the asymptotic regime of the top, namely
for which
value of $\mt ~\, \dro{2}$ begins to be close to the \gmuq\
contribution.
 We found that, typically, $\dro{2}$ starts to be within 10\% the
leading $\mt^4$ value for $\mt \simeq 800$ GeV.

To consider the top as an  asymptotically heavy particle can be an
unrealistic assumption also for  \ew\ quantities of LEP
interest, like
\dlr \cite{Si80} and \drcar \cite{Si89,DFS}. It is then important
 to have  a feeling of how large the theoretical error one is
making  can be  when these quantities are computed including only the
\gmuq\
correction. A possible way to obtain this is to assume that the ratio
between
the \gmud\ and the \gmuq\ contributions in $\dro{2}$ can be
  representative of  the
unknown  two-loop top effects in \dlr\ and \drcar. We can then use
this ratio
to estimate the additional contributions to \dlr\ and \drcar\ simply
multiplying
it by the known \gmuq\ terms of these quantities. The shifts in the
W mass and the effective sinus, $\sineff$,~
due to these additional contributions can be
estimated from the relations
\beqs
\frac{\Delta\, \mw}{\mw} = &~ - \frac{s^2}{2 (c^2 - s^2)}~ \delta
(\dlr) \\
\Delta\, \sineff =&~~ \frac{\scur \ccur}{\ccur - \scur}~ \delta
(\drcar) +
                   \scur \delta \kl (\mz^2)~~~,
\eeqs
where the correction \kl\ is defined in \cite{DS89}.

\begin{table}
\begin{center}
{\bf Table 2}
\end{center}
Calculated ratio $(R)$, for few values of $\mt$ and $\mh$,
        between the \gmud\ and the \gmuq\ contributions in $\dro{2}$.
The corresponding estimate of the shifts in the W mass and $\sineff$
are also presented (see text).
\begin{center}
\begin{tabular}{|c|c|c|c|c|}
\hline
\multicolumn{1}{|c|}{$\mt$} &
\multicolumn{1}{c|}{$\mh$} & \multicolumn{1}{c|}{$R$}
&\multicolumn{1}{c|}{$ \Delta \mw$}
&\multicolumn{1}{c|}{$ \Delta \sineff $}\\
\multicolumn{1}{|c|}{(GeV)} &
\multicolumn{1}{c|}{(GeV)} & \multicolumn{1}{c|}{$\%$}
&\multicolumn{1}{c|}{ (MeV) }
&\multicolumn{1}{c|}{$ (10^{-4}) $}\\
\hline
\hline
    & ~65 & 247 & --10 & 0.6  \\ \cline{2-5}
150 & 250 & 100 & ~--8 & 0.5  \\ \cline{2-5}
    & 800 & ~35 & ~--4 & 0.2  \\ \hline
    & ~65 & 234 & --16 & 0.9  \\ \cline{2-5}
175 & 250 & ~94 & --14 & 0.8  \\ \cline{2-5}
    & 800 & ~38 & ~--8 & 0.5  \\ \hline
    & ~65 & 221 & --23 & 1.4  \\ \cline{2-5}
200 & 250 & ~88 & --20 & 1.2  \\ \cline{2-5}
    & 800 & ~38 & --13 & 0.7  \\ \hline
\end{tabular}
\end{center}
\end{table}

In Table 2  we show,
for few values of $\mt$ and $\mh$, the effect of our estimate of the
unknown
top contributions on the W mass and \sineff. In our estimate we have put
$\delta \kl =0$. The ratio
between subleading and leading terms in $\dro{2}$ has been computed
using
expressions slightly different  from \eqs{e2.14}. In fact,  we decided to
maximize the
one-loop result of our $\msbar$ calculation
by writing it  in terms of the physical masses of both W and Z.
Such a procedure is frequently used in one-loop calculations
\cite{DFS}, and in our case
has the further advantage of eliminating the process-dependent terms.
\ From the third column, it can immediately be seen that,
for a fixed value of the top
mass, the effect is more pronounced for light Higgs. This is not
surprising,
bearing in mind the fact that the \gmuq\ term is a monotonically
increasing (in modulus) function of $\mh$.

We want to  stress that the numbers presented in Table 2, more than a
definite
estimate of the shifts in $\mw$ and \sineff\,  should be taken as an
indication that subleading two-loop $\mt$ effects could be larger than
what is
``na\"\i vely'' expected. Their size is probably comparable to, or
may be
larger than,  the theoretical uncertainty due to the hadronic
contribution to the photonic \self. The latter amounts to $\pm 16 $ MeV
and  $\pm 3 \times 10^{-4}$ in $\mw$ and \sineff, respectively.

To conclude, we think that  our calculation  of $\dro{2}$ shows that
it is
questionable to believe that two-loop electroweak top contributions are
well approximated by the $\gmuq$ term and therefore
sufficiently under control. However, the possibility of establishing top
effects of a purely electroweak nature at the two-loop level seems
quite remote. The experimental accuracy envisaged
for the W mass is $(\delta \mw)_{{\rm exp}} = \pm 50$ MeV, whilst
\sineff\
is presently known with a precision $(\delta~ \sineff)_{{\rm exp}}
\equiv \pm
4 \times 10^{-4}$. At this
level of precision it is likely that only  QCD corrections to one-loop
 top
contribution  can be relevant. However, if  the experimental precision
improves in the future to reach
$(\delta~ \sineff)_{{\rm exp}} = \pm 2 \times 10^{-4}$,
or $\pm 1 \times 10^{-4}$, then a meaningful theoretical interpretation
will
require a complete study of two-loop top effect of electroweak nature.
\subsection*{Acknowledgements}
The authors are indebted to
A.~Sirlin and M.~Tonin for valuable discussions.
One of us (P.G.) would like to thank  the University of Padova for
financial
support, and its Physics Department for its warm hospitality during
his staying in Padua.

\end{document}